\documentclass[vecphys]{svmult}
\usepackage{graphicx}  
\usepackage{subfigure}   
\usepackage{epstopdf} 
\usepackage{mathptmx,helvet,courier}  
\usepackage{multicol}        
\usepackage[bottom]{footmisc} 
\usepackage{amsmath}\usepackage{amssymb}

\usepackage{marvosym}
\begin{document}

\title*{Localized waves in silicates. What do we know from experiments?}
 \titlerunning{Localized waves in silicates. What we know from experiments?}
 \author{F. Michael Russell \and Juan F.R. Archilla \and Santiago Medina-Carrasco}
\authorrunning{F.M. Russell, J.F.R. Archilla, S. Medina-Carrasco}

\institute{  F.M. Russell
  \at School of Computing and Engineering, University of Huddersfield, HD1 3DH, United Kingdom, \email{mika2mike@aol.com}
\and
J.F.R. Archilla (\Letter)
\at Group of Nonlinear Physics, Universidad de Sevilla, ETSII, Avda. Reina Mercedes~s/n, 41012-Sevilla, Spain, \email{archilla@us.es}
 \and
S. Medina-Carrasco
\at X-Ray Laboratory, CITIUS, Universidad de Sevilla, Avda Reina Mercedes 4B,
41012-Sevilla, Spain, \email{sanmedi@us.es}
 } 

\maketitle
\label{chap:archilla}

\abstract{Since the latest review about solitary localized waves in muscovite, called quodons, [FM Russell, Springer Ser. Mater Sci. 221 (2015) 3] there have been many developments, specially from the point of view of experiments, published in several journals.  The breakthrough hypothesis that was advanced in that review that dark tracks were produced by positive electrical charge moving in a localized wave, either transported by swift particles or by nonlinear localized waves, has been confirmed by experiments in muscovite and other silicates. In this paper we review the experimental results, some already published and some new, specially the phenomenon of charge transport without an electric field, called hyperconductivity. We also consider alternative explanations as phase transitions for other tracks. We also attempt  to describe numerical simulations that have confirmed the order of magnitude of quodons energy and calculations underway to determine more properties of electron and hole transport by quodons.
}

\keywords{Layered silicates, nonlinear waves, quodons, kinks, breathers, charge transport, hyperconductivity.}

\section{Introduction}
\label{sec:archilla-introduction}
The existence of localized waves in silicates layers were first proposed in 1994~\cite{collins-russell-lugano1994}.  This was an important step in a long story of research about the nature of tracks in muscovite mica since 1967~\cite{russell67a,russell67b}. A scientific review~\cite{russell-tracks-quodons2015short} and a longer historical review~\cite{russell-crystal-quodons2015short} were published in 2015. The  hypothesis that {\em quodons}, i.e., quasi one-dimensional lattice excitations, transport electric charge was proposed in those reviews but not developed.

This hypothesis was a fundamental change that led to new theory, new interpretation of previous  results about tracks in muscovite, and specially to experiments that confirmed and modified the theory. Therefore, we have thought that it was time for a new review that provides a comprehensive and brief summary of the state of knowledge and the challenges in front of the research.

The research can be divided in three stages that are interconnected.
\begin{enumerate}
\item {\bf Tracks}: Tracks by swift particles.
\item {\bf Quodons}: Tracks by lattice excitations or {\em quodons}.
\item {\bf Hyperconductivity}: Quodons with electric charge and {\em hyperconductivity}.
\end{enumerate}

Here we present the beginning and end of the three stages and some of the highlights. Later, we will explain in detail some key aspects.

\subsection{{\em Tracks} by swift particles}
This stage starts in 1967 with the observation in mica of dark tracks of charged particles
from neutrino interactions~\cite{russell67a} and finishes in 1993 with an explanation of track formation  by release of lattice energy~\cite{russell-release1993} and the description of semi-transparent tracks in mica related with positron dark tracks~\cite{steeds-russell1993}. Dark tracks are made out of magnetite and some shorter semi-transparent tracks are made out of the mineral epidote.

Note that tracks are also the result of experiments similar to particle tracks in a bubble chamber. They are experiments that nature has made and have been conserved as a fossil in muscovite crystals. They have been done at temperatures, pressure and specially time scales outside of the possibilities of physicists.

\subsection{Tracks by lattice excitations or {\em quodons}}
From the very beginning~\cite{russell67a} it has been observed that only
0.1\% of the dark tracks in muscovite were produced by swift particles, while the rest lie along the close-packed direction within the cation layers and therefore are related with the crystal structure. This second stage starts with the  calculation of nonlinear forces between potassium ions and using them to obtain an approximate KdV equation for lattice displacements. The KdV equation supports soliton solutions~\cite{KdV1895}, therefore the majority dark lines in muscovite could be produced by lattice-solitons. These results were presented at a conference in 1994~\cite{collins-russell-lugano1994} and extended the following year~\cite{russell-collins95a,russell-collins95b}.

Interestingly, in the same year 1994, it was attempted to observe lattice-solitons by  bombarding silicon with 0.8\,MeV Ar$^+$ and detecting the ejection of an atom~\cite{schlosser1994}. The experiment failed, perhaps among other reasons because it used silicon which is not layered and have a complicated structure for soliton propagation as the nearest neighbours do not form straight lines.

These lattice excitations were named {\em qodons} in 1995~\cite{russell-collins95b} and later {\em quodons} in 1998~\cite{marin-eilbeck-russell-2Dhexa1998}. This was an acronym for quasi one-dimensional excitations, a descriptive term which also recognized that the actual type of excitation was not well known. It is worth noting that  the term lattice-soliton was changed to {\em breather}. Breathers differ from solitons in having an internal vibration and smaller energy and were starting to be thoroughly studied~\cite{mackayaubry94,flach1995}.

The highlight of this stage is probably the success of another experiment in 2007~\cite{russell-experiment2007} similar in design to the previous one~\cite{schlosser1994}. In this case a mica monocrystal was bombarded with alpha particles and it was possible to detect the ejection of atoms at the opposite side of the sample along the direction of close-packed lines within the cation layers.

This stage finishes in 2015 with two comprehensive reviews, a shorter and scientifically oriented one~\cite{russell-tracks-quodons2015short} and a longer historical review oriented to the non-specialist~\cite{russell-crystal-quodons2015short}. But in these two reviews the next stage is also hinted.

\subsection{Quodons with electric charge and {\em hyperconductivity}}
It was well known that most tracks in muscovite were produced by the recoil of potassium atoms after beta decay~\cite{russell88-identification}. In 2015, a thorough analysis of the decay modes of $^{40}$K~\cite{archilla-kosevich-quodons2015,cameron2004,radionuclides2012} showed that 90\% of decays left a charge behind, and this  charge was positive except in 0.001\% of positron decays, when it was negative. Then, it was realized that dark tracks by swift particles were produced only by positive particles and that the thickness of, for example, positron tracks, at sonic speed, when they were about to stop, were similar to quodon tracks. These two observations led to the deduction that quodons have electric charge, and dark tracks, positive charge~\cite{russell-arxiv2015}. This hypothesis was already introduced at the previous reviews and it was later extended in Ref. \cite{archillaLoM2016}.

This profound change in the quodon concept provided something to measure easily, electric current, when quodons were excited by particle bombardment without an electric field, a phenomenon called {\em hyperconductivity}. Experiments were successful and also were able to explain new properties of quodons~\cite{russell-archilla2017,russell2019} in muscovite and other layered silicates.

\begin{figure}[t]
\begin{center}
\sidecaption[t]
\includegraphics[width=9cm]{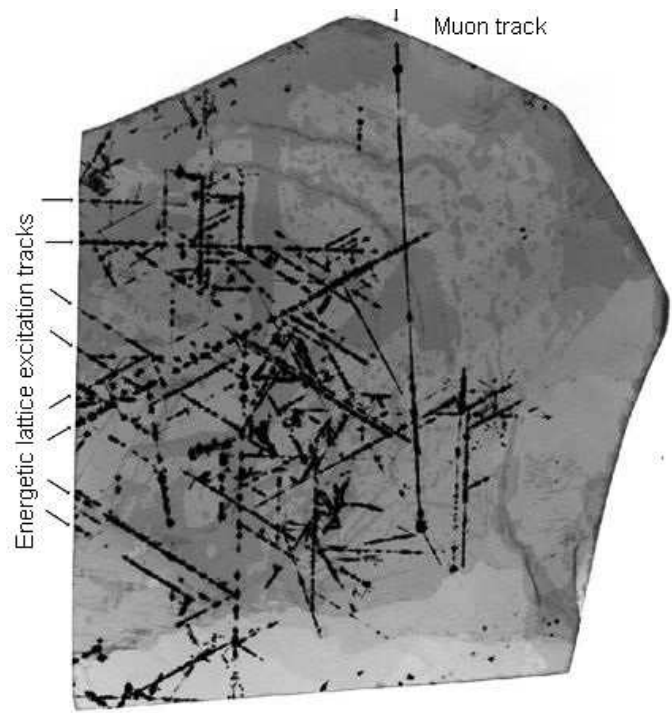}
\caption{A sheet of mica muscovite showing many majority tracks due to lattice excitations within the hexagonal structure of the cation layer and a muon track in an unrelated direction}
\label{figure_micasheet}
\end{center}\end{figure}

\section{Important points}
In this section we concentrate in some important points which illustrate either fossil tracks or experimental results or experiment setup.
\subsection{First encounter with dark tracks in muscovite}
It is important to emphasize that the main author of this research F.M. Russell has been all his career dedicated to high energy physics, first at Harwell Laboratory\footnote{Atomic Energy Research Establishment near Harwell, Oxfordshire, U.K.}, then at Oak Ridge National Laboratory (ORNL) in the U.S.A, and thereafter at the Rutherford Appleton Laboratory (RAL) in the U.K. In this way, when in 1963 at a museum in North Caroline\footnote{Museum of North Carolina Minerals, Spruce Pine, North Caroline, U.S.A.}, he found himself in front of a specimen of muscovite  with abundant dark tracks, he recognized the striking similarity with the tracks of swift particles in bubble chambers. A similar sheet is presented in Fig.~\ref{figure_micasheet}.

\begin{figure}[b]
\begin{center}
\sidecaption[t]
\includegraphics[width=7.5cm]{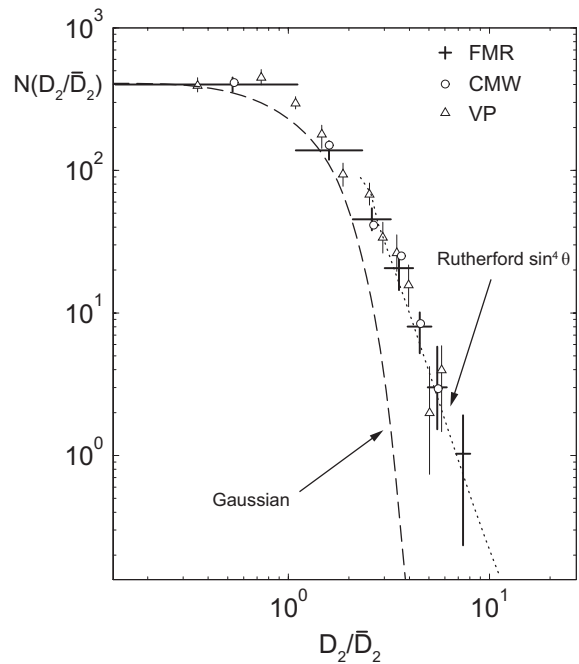}
\caption{Probability of scattering at given angles for tracks corresponding to positrons
in muscovite [+], compared with
positrons in a photographic film [$\triangle$] and by Wolfendale's group [$\circ$].
The results fit
closely to the Rutherford Law, thus strongly supporting the
hypothesis that the lines are tracks
of charged particles. Data from 
Refs.~\cite{voyvodic1952,wolfendale68,russell88-identification}
for VP, CMW and FMR, respectively. Reproduced with permission
from: Russell, F.~M. (1988)\cite{russell88-identification}
 Copyright $\copyright$ 1988, Elsevier}
\label{figure_kinkiness}
\end{center}\end{figure}

\subsection{How were the swift particles identified?}
There were different methods, but perhaps the clearest is the kinkiness of those dark tracks. Charged swift particles when entering in matter experience scattering with the matter ions. The probability of scattering at a given angle can be calculated by Rutherford law and the angles can be seen and measured with a microscope and the results compared with given particles. An example can be seen in Fig.~\ref{figure_kinkiness}, comparing the second difference, basically the scattering angle, of some track in mica with positrons in a photographic film~\cite{voyvodic1952}, taking into account the difference in mass and density of the scattering ions.

\subsection{Which particle tracks were identified?}
The particles that produce dark tracks in muscovite and could be identified were positive muons, i.e., antimuons, which are the particles that can be produced deep underground after neutrino interaction~\cite{russell67a,russell67b,russell68,russell88-identification},   positrons from $^{40}$K decay and  antimuon decay~\cite{russell88-identification,russell88-positive,russell-decorated91a,russell-channelled-91b,steeds-russell1993}. Protons can be recognized by the short length of the tracks corresponding to non-relativistic speed~\cite{russell-persistent2011,russell-tracks-quodons2015short}. Also, alpha particles can be discriminated from the multiple scattering events, proof of their large energy and mass~\cite{russell68,russell-tracks-quodons2015short}.

The remarkable fact that all the particles that produce dark tracks were positive was used in 2015 to recognize that the large majority of quodons that produce dark tracks have also positive charge~\cite{russell-arxiv2015,russell-tracks-quodons2015short,russell-crystal-quodons2015short}.

\subsection{How were the tracks produced?}
There is not enough energy to produce the dark tracks, this means that the source of energy is already in the lattice, in the form of a metastable state~\cite{russell-decorated91a,russell-channelled-91b,schlosser1994,russell-collins95b}.

Natural crystals of muscovite mica contain various impurities, especially iron, incorporated during their growth. It has been found that this can lead to a unique situation, as a crystal cools following growth, during which minute perturbations of the crystal can be recorded and stored indefinitely. Although muscovite is a common mineral in rocks, large crystals grow only in pegmatites associated with magmas at temperatures of about 500$^\circ$C and under high pressure at about 5\,km underground~\cite{deer2013}. Inevitably, large single crystals of good quality are rare but they are of special interest because of the information they have been found to contain. A common feature of micas is their ease of cleavage, in the (001)-plane.
The black material forming the patterns is the iron oxide mineral magnetite, so named because it is ferro-magnetic.

As a crystal cools slowly at high temperature it tries to reach a lower energy state by expelling the magnetite at the weakest part of the lattice, the cleavage plane. The magnetite grows epitaxially, centred in the potassium sheet and grows in the directions of structural weakness. These are the principal crystallographic directions, which are easily determined by percussion figures~\cite{russell-crystal-quodons2015short}. This has been confirmed by both optical and electron microscopy. In fact, the distortion of the lattice is readily seen by observing the of the intrusive magnetite by reflected light or by surface interferometry. Contrary to the basic assumption in of global bi-stability of structure there is no evidence for this in the observed patterns involving magnetite. 

\subsection{Two different recording processes}
It has been found that there are two different recording processes leading to the observed patterns, involving different impurities. The dominant process leading to magnetite is triggered by passage through the crystal of a positive charge in the vicinity of the potassium sheets. This can result from a positively charged, high-energy, muon created in a neutrino interaction within the Earth or by direct penetration of a cosmic ray. Another source is from electron-positron showers arising from a high-energy gamma interaction. The flight-paths of these particles are influenced by channelling and diffraction scattering due to the pronounced layered structure of muscovite~\cite{russell88-identification}. The most informative source, however, is from the rare decay channel of $^{40}$K creating positrons~\cite{steeds-russell1993}. Study of the fossil tracks of these positrons has shown that the origin of the nucleation sites for triggering magnetite growth does not involve ionization of the lattice.  For relativistic positrons from this source a fossil track results even when the rate of energy loss is less than 1\,eV per 10,000 atoms along the flight path. The rate of energy loss increases as a positron slows down, leading to an increase in the amount of magnetite formed. Due to anisotropy of the mechanical properties of the layered structure this increase shows as a widening of the magnetite ribbon delineating the flight-path. This suggests that the recording process is of a chemical nature, with the probability for an impurity ion migrating to the flight path increasing as the positron's speed decreases.The dominant source of the long ribbons of magnetite arising from moving positive charges is the dominant decay channel of $^{40}$K, in which an electron is emitted. These energetic electrons do not initiate fossil magnetite tracks. However, they leave a positive charge at the decay site that can be trapped and carried by a mobile lattice excitation arising from the recoil motion of the decayed nucleus. These mobile, non-dissipative, highly localized excitations move at slightly sub-sonic speed, leading to magnetite ribbons of width of similar width to those due to nearly stopped positrons~\cite{russell-tracks-quodons2015short}. The last known source of swift positively-charged ions is from atomic cascades arising from nuclear scattering of relativistic particles.
\begin{figure}[t]
\begin{center}
\includegraphics[width=5.6cm]{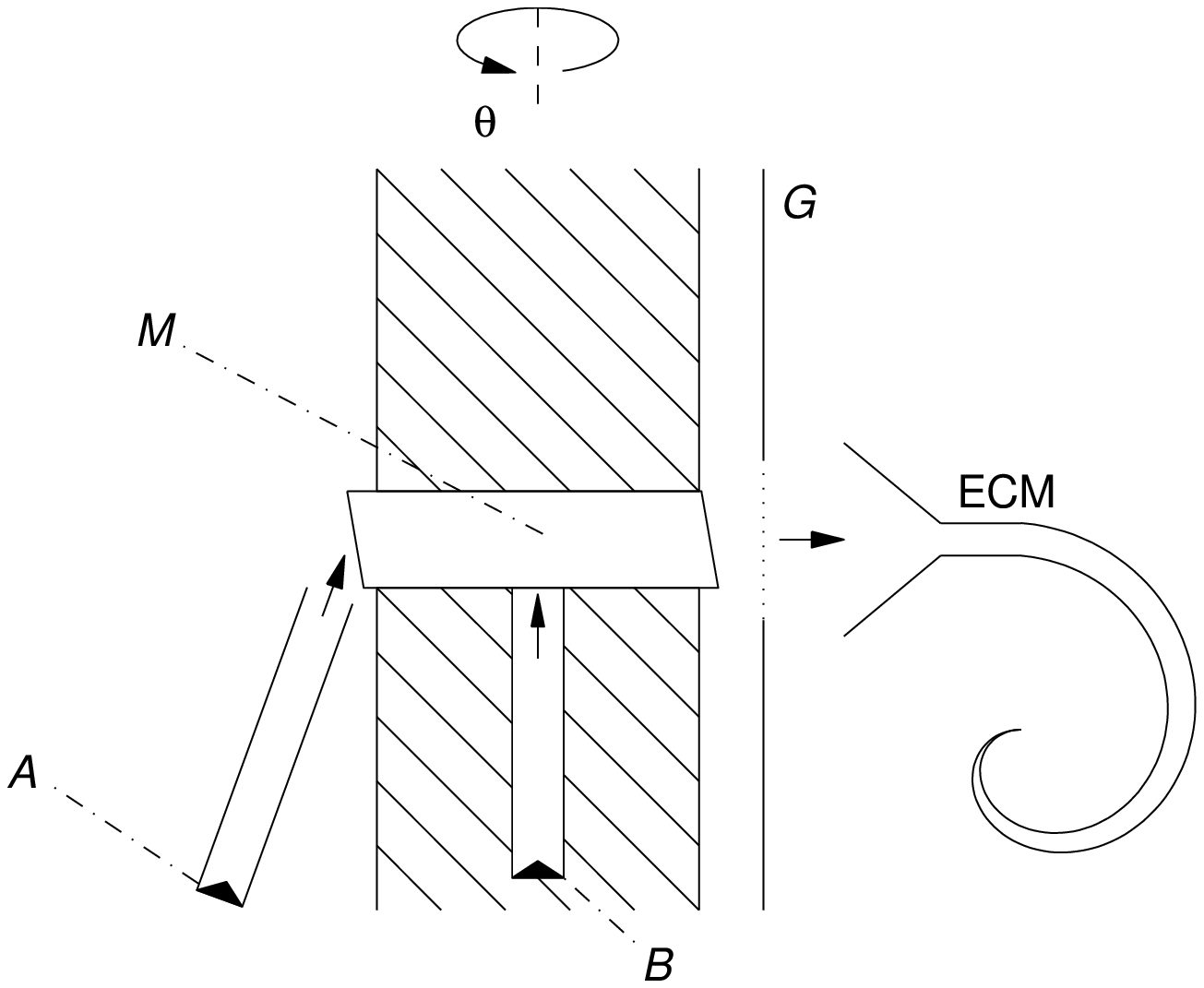}
\includegraphics[width=6.0cm]{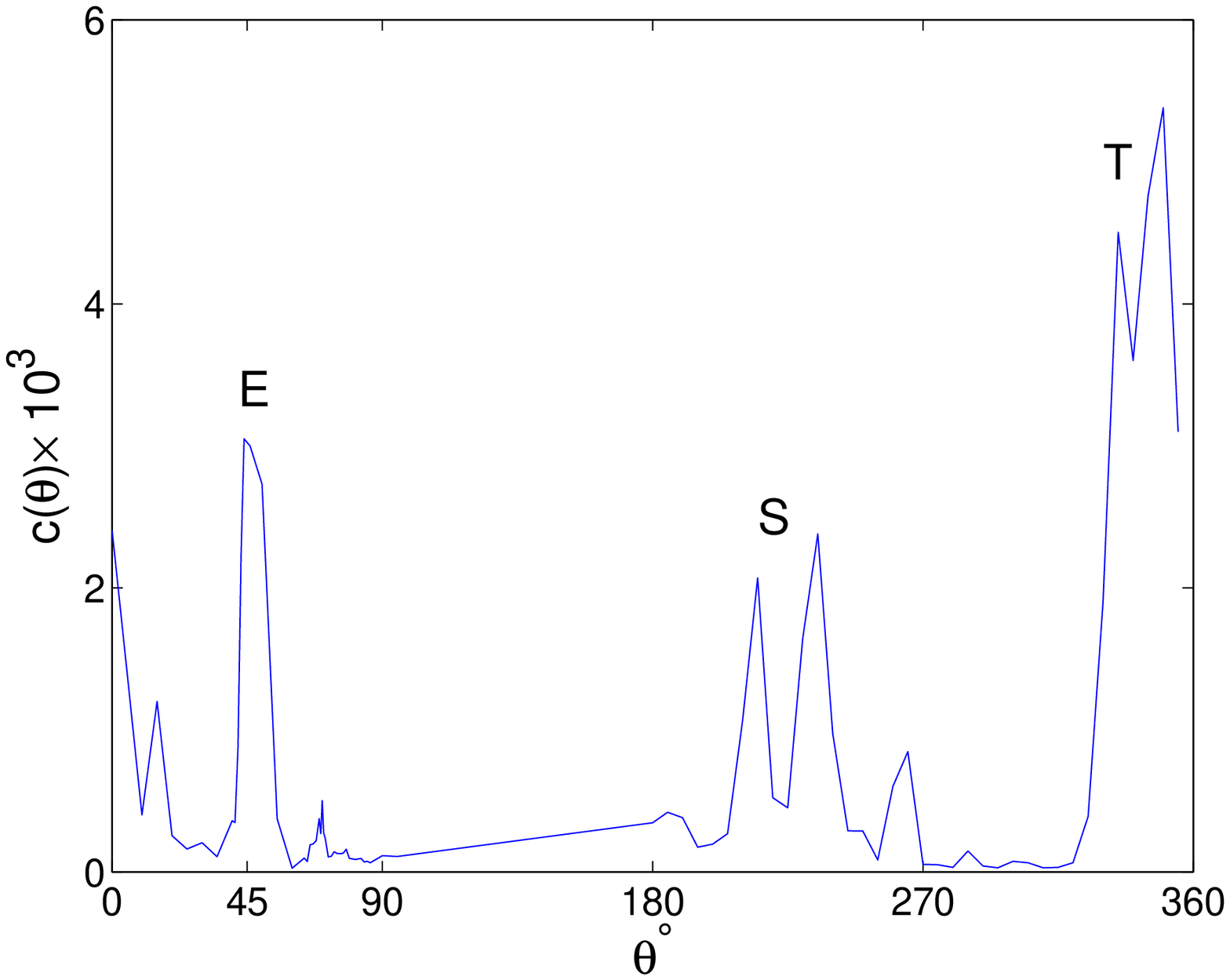}
\caption{{\bf Left}: Setup of the quodon experiment. A: alpha source, M: mica monocrystal, ECM: electron channel multiplier, G: grid, B: alternative position for the alpha source.
{\bf Right}: Outcome of the quodon experiment: Plot of the angular dependence of the ECM count rate. T: test, S: sputtering from the front face, E: peak from ejected atoms at the rear face in the [0 1 0] direction.
Reproduced with permission
from: Russell, F.M. and Eilbeck, J.C.~\cite{russell-experiment2007}. Copyright $\copyright$ 2007, EPLA.
}
\label{figure_quodonexperiment}
\end{center}\end{figure}

The second and much rarer recording process involves formation of the mineral epidote, which requires an excess of calcium during crystal growth. These fossil tracks arise from the emission of a positron, leaving a negative charge at the decay site, which is trapped and transported by the mobile recoil excitation. This leads to a ribbon of transparent epidote that is not intrusive in the potassium sheets~\cite{steeds-russell1993}. The formation process of the epidote is poorly understood and might involve a bi-stable crystal state~\cite{krylova2020}. It is hoped that this explanation of the origin of the fossil magnetite-ribbon tracks might encourage study of the formative process for the fossil epidote tracks, as this has the potential for ballistic, low-loss, transport of electrons in layered insulators\, 
\cite{russell-archilla2017,steeds-russell1993,russell2018,russell2019}.

\section{How was the experiment in lattice-excitations or quodons done?}
\label{sec:quodon experiment}
The highlight of the research on lattice excitations, sometimes called lattice-solitons, breathers or quodons in this context was the experiment in 2007~\cite{russell-experiment2007}. Alpha particles were sent at an angle with the muscovite sheet and therefore with the potassium layer to prevent the possibility of transmission and it was detected at the other side of the monocrystal corresponding to low Miller indexes, the ejection of an atom from the surface. The atom was detected because it was ionized by an electric field and the charge detected.  Ejection of atoms from a silicate surface needs energies of 7-8\,eV, however, it is not necessary that a quodon has that energy as the passage of a vibrational energy in the vicinity of the surface is enough to increase the probability of ejection~\cite{dubinko-archilla2011}.  Both the setup and the outcome can be seen in Fig.\,\ref{figure_quodonexperiment}.

\begin{figure}[t]
\begin{center}
\includegraphics[width=5.8cm]{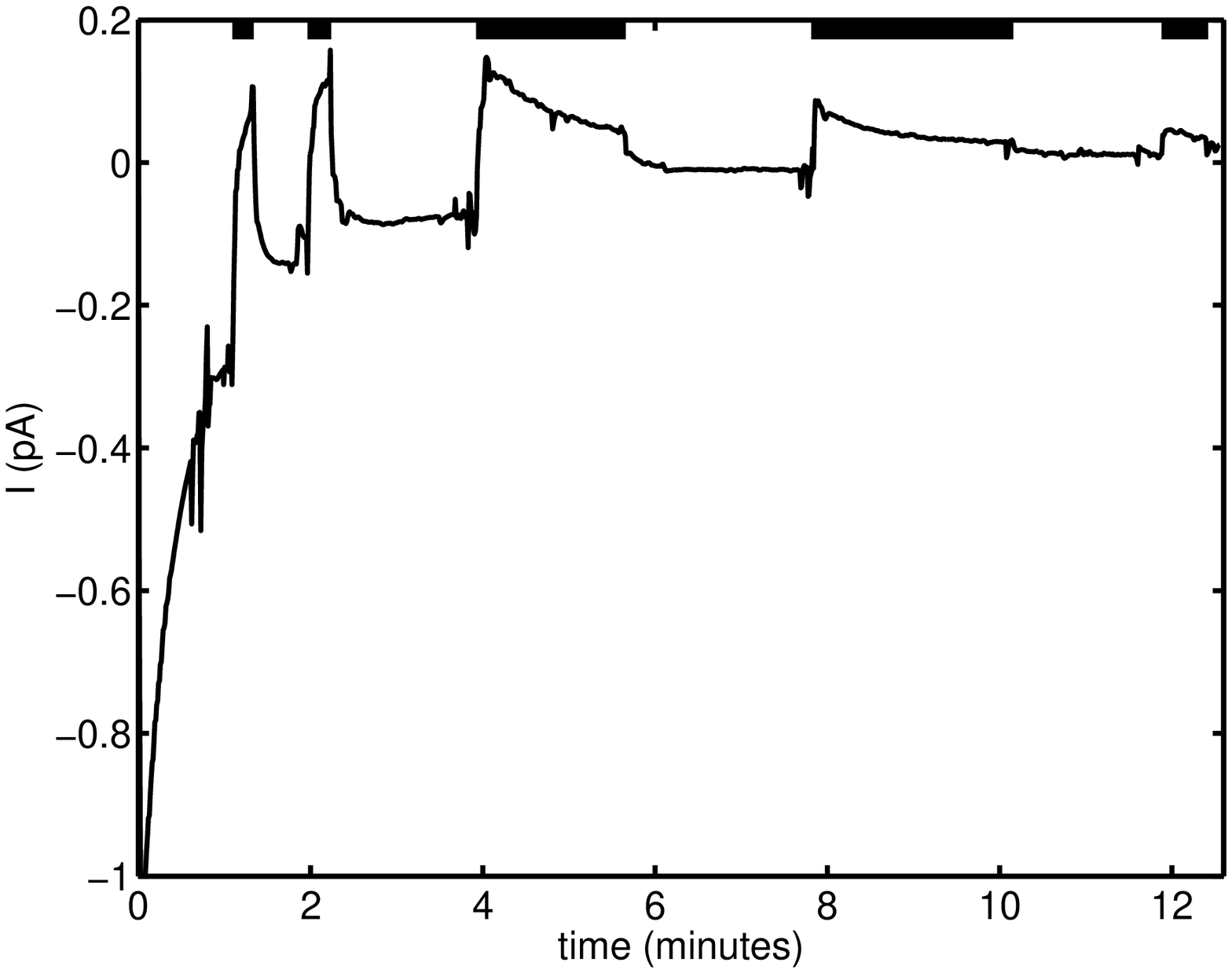}
\includegraphics[width=5.8cm]{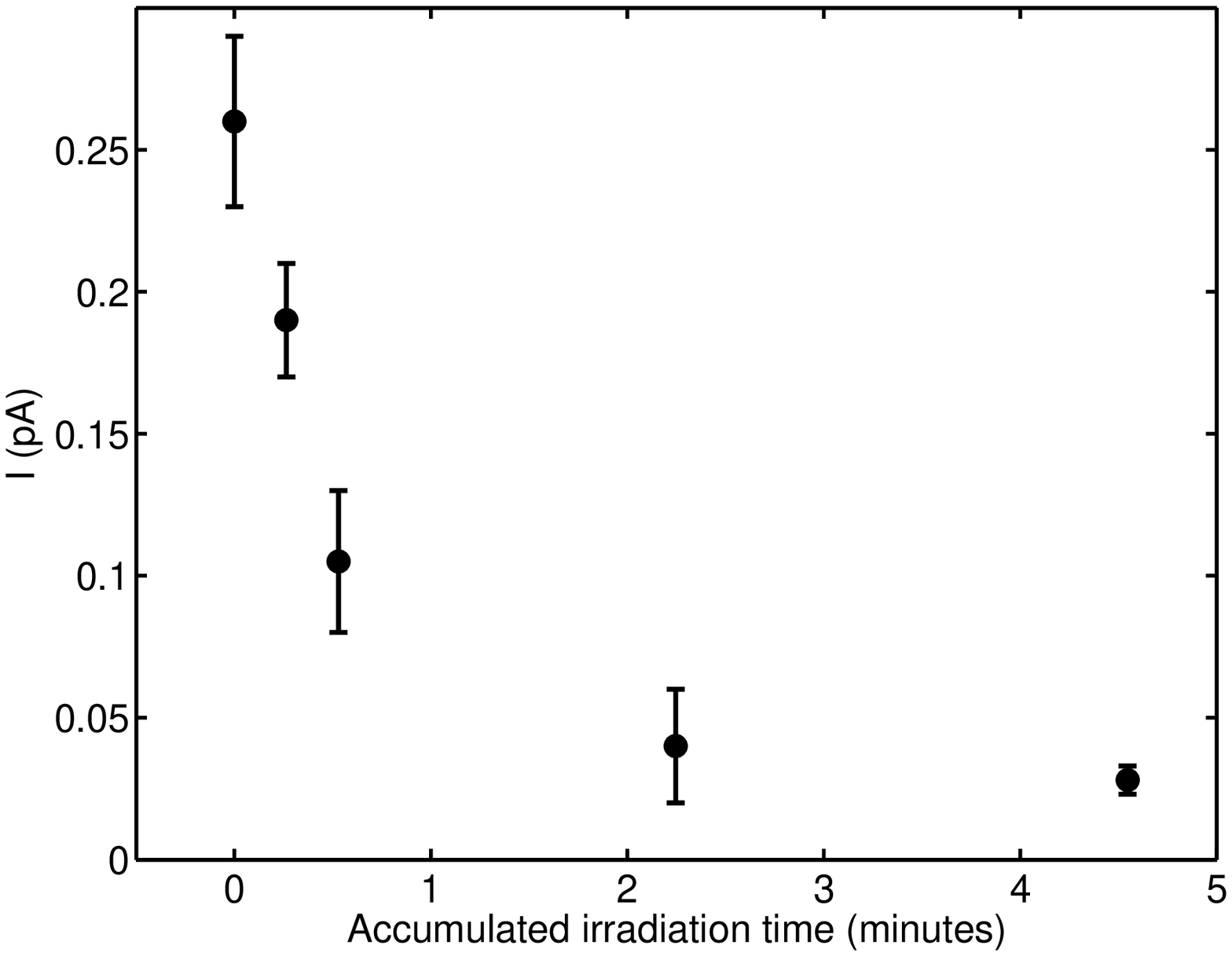}
\caption{{\bf Left}: Plot of the hyperconductivity current, the time intervals marked in black at the top correspond to the opening of the alpha gate.
{\bf Right}:  Hypercurrent as a function of the accumulated time of alpha exposure, showing an exponential decrease corresponding to the depletion of the charge reservoir from $^{40}$K beta decay.
Reproduced with permission
from: Russell, F.M.  et al.~\cite{russell-archilla2017}. Copyright $\copyright$ 2017, EPLA.
}
\label{figure_hyper1}
\end{center}\end{figure}

\section{How it was demonstrated hyperconductivity, i.e., that quodons carry charge?}
Hyperconductivity is defined as the transport of charge in absence of an electric field. The charge is transported by nonlinear excitations which have their own energy and momentum from the cause that created them. Due to the combination of nonlinearity and discreteness they travel long distances in atomic terms with little attenuation.

An experiment was set up quite similarly to the previous ones. The way to excite lattice excitations or quodons was also by sending alpha particles, due to its simplicity. The hypothesis was that alpha particles would produce many quodons and some proportion of them would propagate to the other side of the sample and in this way a current could be measured. Muscovite is a very good insulator but there was the possibility that the surface and certainly the ionized air would transport charge. To discard this effect the two sides of the sample were connected and therefore the potential difference among both contacts would be zero and also the electric field would be zero. Lattice excitation or quodons would travel due to their initial energy and momentum.

The experiment was a success but with some unexpected results. Instead of a steady current after the alpha gate was open, the current showed a peak, but then it would diminish to a small limiting value. The phenomenon was soon explained: there are not free carriers in muscovite band structure, the available charge is the one obtained after beta decay of $^{40}$K, mainly positive after $\beta^-$, i.e., the emission of an electron is the dominant branch, but also some negative charge after $\beta^+$ positron emission. This reservoir is depleted in some minutes, and the remaining current is exactly the flux of electric charge brought by the alpha flux~\cite{russell-archilla2017}. The current peaks and their decrease can be seen in Fig.~\ref{figure_hyper1}.

\section{What properties of hyperconductivity and quodons were deduced from experiments?}
\begin{figure}[t]
\begin{center}
\includegraphics[width=5.8cm]{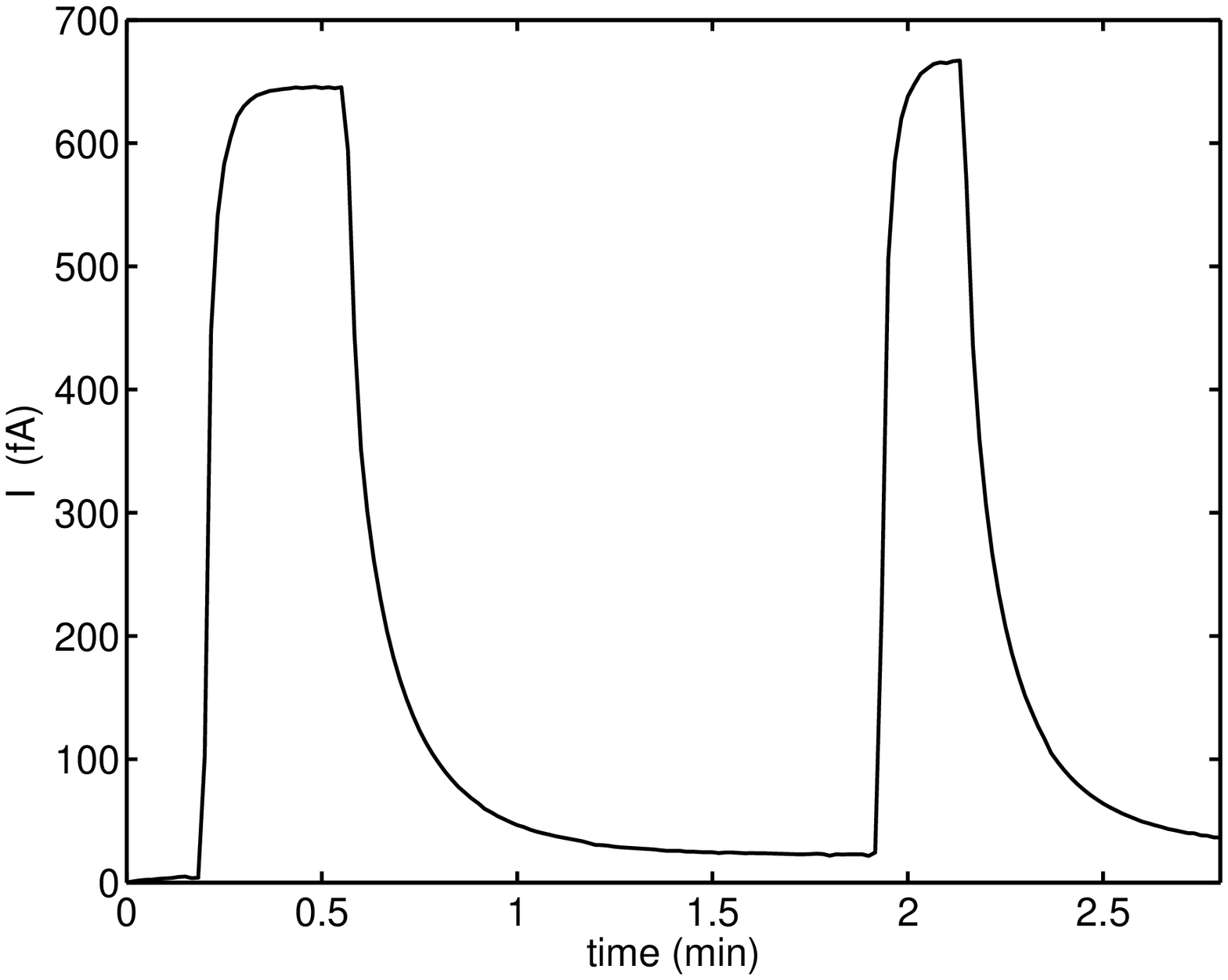}
\includegraphics[width=5.8cm]{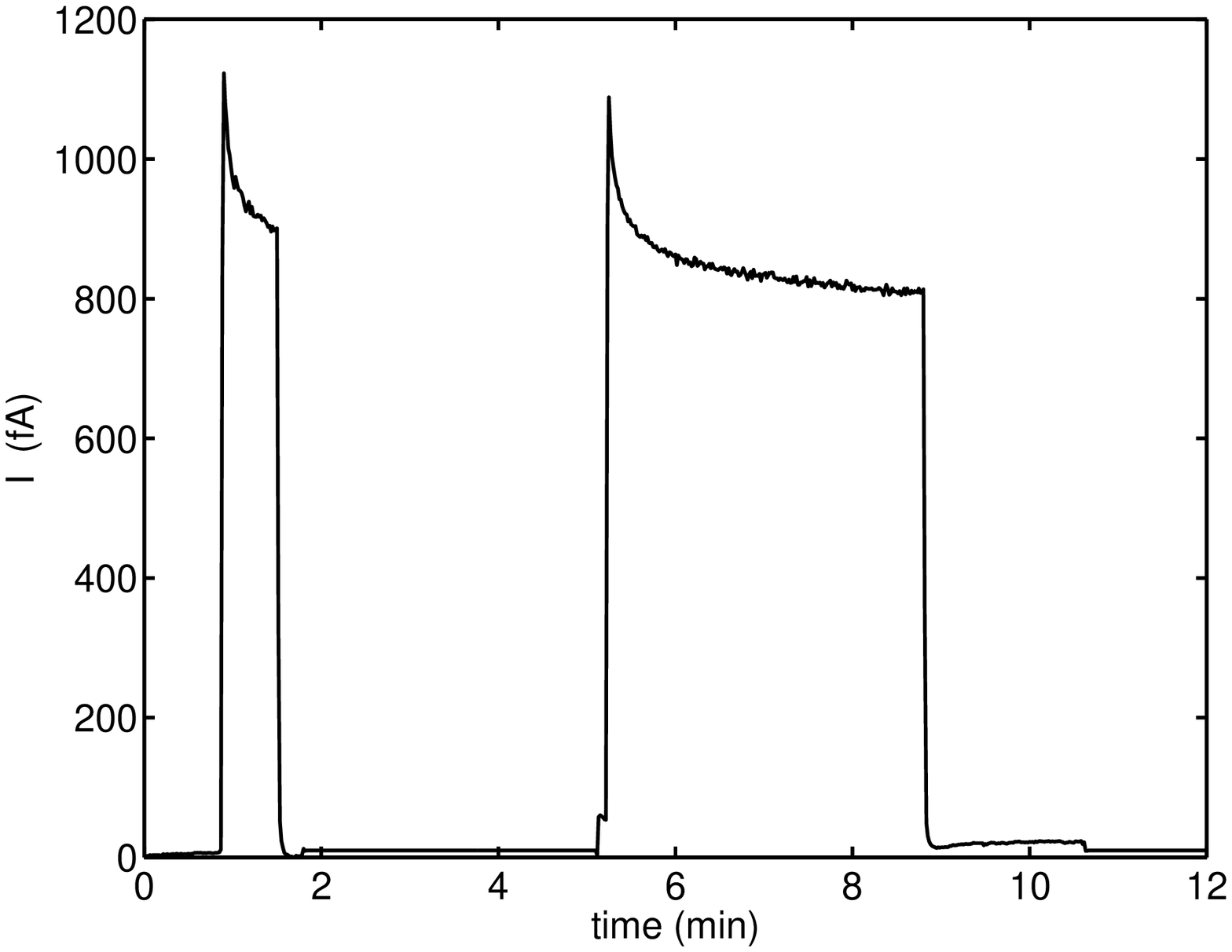}
\caption{Plot of the hypercurrent corresponding to two intervals of alpha exposure in a previously depleted crystal. {\bf Left}: Crystal of lepidolite of good quality. Note the soft decay of the hypercurrent after the alpha flow is stopped.
{\bf Right}:  Crystal of phlogopite of bad quality. Note the abrupt decrease of the hypercurrent after alpha irradiation is stopped.
Reproduced with permission
from: Russell, F.M.  et al.~\cite{russell2019}. Copyright $\copyright$ 2019, EPLA.
}
\label{figure_flightpath}
\end{center}\end{figure}

More experiments in hyperconductivity~\cite{russell2019} were able to deduce a number of facts:
\begin{itemize}
\item Other layered silicates as lepidolite, phlogopite, chrysolite and both natural and synthetic fluorphlogopite supported hyperconductivity and thus the propagation of quodons. However, a layered silicate as biotite with similar structure does not support it. It was not found in unrelated materials that could be used in quodon technology as PTFE, quartz, borosilicate glass and epoxy resin.
    \item Hyperconductivity is not sensitive to minor crystal defects and can even anneal some of them. It can also pass through some interfaces.
    \item Hyperconductivity is not affected by magnetic fields up to 1.1\,T.
    \item Quodons have very long flight paths, this can be deduced by comparing the drop in hypercurrent when the alpha bombardment is stopped. In a good crystal the hypercurrent continues to flow some seconds, while in a crystal with many defects, the hypercurrent stops almost immediately as can be seen in Fig.~\ref{figure_flightpath}
\end{itemize}

\section{What types of quodons are there?}
\begin{figure}[t]
\begin{center}
\includegraphics[width=\textwidth]{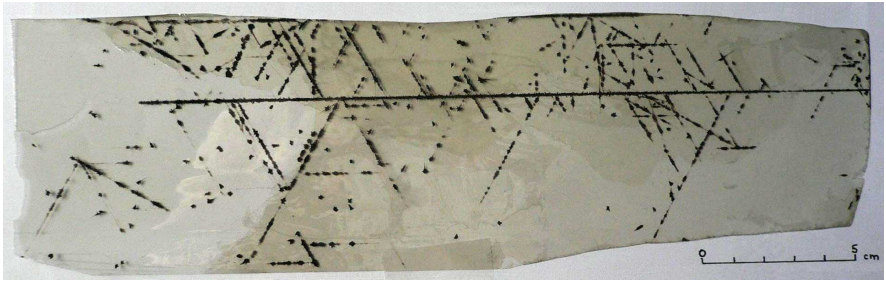}
\caption{A sheet of mica muscovite showing a quodon primary track and many secondary tracks scattered from it. Also it is possible to see the intermittency in the secondary tracks along the close packed direction of the cation layer. This  is interpreted as a quodon loosing and regaining a positive charge.
Reproduced with permission
from: Russell, F.M.~\cite{russell-crystal-quodons2015short}. Copyright $\copyright$ 2015, Springer.
}
\label{figure_primary}
\end{center}\end{figure}

There is no clear information from the experiments, however from the fossil tracks as seen in Fig.~\ref{figure_primary}, it can be deduced:
\begin{itemize}
\item There are positive quodons, negative quodons and probably neutral quodons. Negative quodons can be seen as an epidote track in exactly the opposite direction from a positron track and therefore corresponding to the recoil of the nucleus of $^{40}$K after $\beta^+$ decay.  Neutral quodons can be deduced from intermittent dark tracks, which seem like quodons loosing an regaining positive charge.
\item There are some more energetic quodons that produce straighter and thicker dark tracks and some less energetic quodons because they appear often as weaker dark tracks scattered from a primary track. As both types are dark, it is deduced that both have positive charge. They might have different nature, maybe primary tracks could be crowdions or kinks as they transport charge in an ionic crystal and have large energies of 20-30\,eV~\cite{dou2011,archilla-kosevich-quodons2015,bajars-quodons2015,archilla-kosevich-pre2015,archilla-zolotaryuk2018}. Secondary tracks, could be interpreted as breathers, because they have good mobility in mica models with little or no radiation, with energies of 0.2-0.3\,eV~\cite{marin-eilbeck-russell-2Dhexa1998,bajars-physicad2015,archilla-osaka19} and recently they have been shown to scatter in different close-packed directions~\cite{bajars2020}. However, breathers do not transport charge and if they couple to a charge their properties and physical description change completely. Certainly, breathers could correspond to neutral quodons. A model for lattice excitations  coupled to a hole or electron has been constructed for muscovite, but the properties of localized excitations using it are still under study~\cite{archilla-nolta2019,archilla2020}.
\end{itemize}

\section{Alternative explanations of tracks}
There have not been many alternative explanations of tracks in muscovite. It was suggested that the majority of dark lines corresponds to dislocations because they lie in the close-packed directions, but without further proof~\cite{wolfendale68}. Arguments against dislocations are that they should appear along crystal fractures, which  does not occur~\cite{russell88-identification} and that dark tracks do not continue to the edge of the crystal specimen as it should occur with dislocations~\cite{russell-tracks-quodons2015short}. Recently an interesting explanation based on phase transition in a bistable lattice has been proposed~\cite{krylova2020}. The research was based on the observation\footnote{J.F.R. Archilla, private communication (2019).} 
 that the pitch of the on-site potential and the equilibrium distance of the interatomic potential should be different in a real material. This bring about the existence of different stable configurations, and the authors found a switching wave between configurations  that propagates longitudinally along the direction of atomic chains. They used a Frenkel-Kontorova 2D system with Morse interaction potential. There was no attempt to relate their findings with physical magnitudes and to explain the coloration of lines or the kinkiness of the swift particle tracks. Also, the hyperconductivity experiments were not explained and the charge of the ions in the cation layer were not taken into account as explained in the article. Nevertheless, it opens a new path to understand some of the phenomena observed in muscovite  and other layered silicates, particulary epidote tracks, which are not produced by swift particles.

\section{Summary}
\label{sec:summary}
In this article we have tried to present an updated review of the research in nonlinear waves in layered silicates, particularly, but not only, in mica muscovite. We have attempted to make clear for the non specialist which are the main experimental facts and their interpretation, leaving many details to the references. The main results are that some dark tracks in muscovite can be related to swift positive particles, that many other tracks along atom chain directions of the cation layers can be interpreted as lattice excitations, called quodons.  Most quodons carry positive charge although some may have negative charge or none. This was demonstrated by hyperconductivity experiments, that is, the transport of charge in the absence of an electric field.  Variants of hyperconductivity experiments allowed for the deduction of many properties of quodons. Other interpretations of dark tracks may be complementary and be useful to understand some of the tracks.


\section*{Acknowledgments}
JFRA thanks a travel grant from VIPPITUS 2020 and projects PAIDI 2019/FQM-280 and MICINN PID2019-109175GB-C22.
SM-C acknowledges grant from PPITUS-2018.

\newcommand{\noopsort}[1]{} \newcommand{\printfirst}[2]{#1}
  \newcommand{\singleletter}[1]{#1} \newcommand{\switchargs}[2]{#2#1}

\end{document}